# *Static-to-dynamic field conversion with time-varying media*

Mario Junior Mencagli[(1)], Dimitrios L. Sounas[(2)], Mathias Fink[(3)], and Nader Engheta[(4)]

Department of Electrical and Computer Engineering, University of North Carolina, Charlotte, NC 28223, USA
Department of Electrical and Computer Engineering, Wayne State University, Detroit, MI 48202, USA
Institut Langevin, ESPCI Paris, Paris, 75005, France
Department of Electrical and Systems Engineering, University of Pennsylvania, Philadelphia, PA 19104, USA

mmencagl@uncc.edu, dsounas@wayne.edu, mathias.fink@espci.fr, engheta@seas.upenn.edu

**Abstract**

In this Letter, we theoretically demonstrate that a uniform static electric field distribution can be partially converted to radiation fields when a portion of the medium undergoes a temporal change of its permittivity. An in-depth theoretical investigation of this phenomenon is developed for a dielectric block with a step-like temporal change located inside a waveguide charged with a DC voltage source. Closed analytical expressions are derived for the radiated electric and magnetic fields. The exchange of energy between the electrostatic and electromagnetic fields is discussed. The reconciliation between the seemingly contradictory temporal and spatial boundary conditions for the electric and magnetic fields at the interface of the time-varying dielectric block is analyzed and elucidated. Our findings may provide an alternative solution for generating electromagnetic radiation based on time-varying media.

Investigations on the electrodynamics of time-varying media, which possess temporal dependent electromagnetic parameters (permittivity and/or permeability), date back to the mid part of the last century. The first investigation belongs to Morgenthaler [1], who published pioneering work on dynamics of plane waves propagating in an unbounded homogeneous medium with a temporal dependent refractive index. It was demonstrated that an abrupt temporal change in the medium refractive index leads to a backward- and a forward-propagating wave, showing the temporal analogue of the spatial interface between two media with different electromagnetic parameters. Since then, other aspects of wave propagation in time-varying media, including source-dependent phenomena [2] and reflection (transmission) from (into) semi-infinite temporal slab [3], have been explored [4]. In recent years there has been an increasing interest in time-varying media from both the engineering and the physics communities. This growth of interest is driven primarily by the fact that they have opened the door to various interesting wave phenomena such as nonreciprocal transmission [5], virtual absorption [6], [7], and temporal aiming [8], to name a few.

One particularly interesting effect in time-modulated media is conversion of static energy to radiation. In the past decades, this phenomenon, aimed at exploring alternative sources of electromagnetic radiation, has been investigated in the context of plasma physics [9]–[13]. In the pioneering work [9], it has been shown how the interaction between a gas biased with a periodic static electric field and a plasma ionization front gives rise to electromagnetic radiation. This effect has been proven very useful for the design of high-power sources showing the importance of media with step-wise modulation. Demonstrating similar effects in dielectric media may open exciting opportunities for new types of sources, antennas, and energy converters.

Here, we analyze a related phenomenon in time-varying dielectric media, which is the main form of time-varying media nowadays. We demonstrate that applying an abrupt change in the electric permittivity of a dielectric slab immersed in an electrostatic field leads to controllable conversion of electrostatic energy to radiation. We investigate this effect through an in-depth theoretical analysis backed up by computer simulations. Finally, we discuss the roles of the continuity conditions of the electric and magnetic fields' tangential components at the interface of a stationary (no time-varying) medium and a time dependent medium, and show how they can be reconciled, overcoming a contradiction that seems to exist between them.

A conceptual representation of a time-varying dielectric structure for converting electrostatic to radiation fields is displayed in Fig. 1. Let us assume a dielectric block with permittivity $\varepsilon_1$ is immersed in a uniform static electric field ($\mathbf{e}_0$), as shown in Fig. 1(a). Then, we assume that the dielectric block, after it has been fully polarized by $\mathbf{e}_0$ and the system has reached the steady state, undergoes a change of its permittivity in time in a rapid step fashion from $\varepsilon_1$ to $\varepsilon_2$, as shown in Fig. 1(b). The permittivity is assumed to be dispersionless, and both values are taken to be real positive larger than unity. The dispersionless assumption is justified when the resonance frequency of the material is very high [14]. This change of the permittivity perturbs the electrostatic field distribution inside the dielectric block established by $\mathbf{e}_0$. Such perturbation results from the condition of temporal continuity of the electric displacement $\mathbf{d}$ [15] at the time instant ($t = t_0$) of changing the permittivity value. However, the material parameters of the region outside the dielectric are assumed to be temporally stationary, and thus no temporal boundary conditions are applied there. It might appear that the tangential components of the electrostatic field are discontinuous across the spatial boundary of the dielectric block in Fig. 1(a) when its permittivity experiences a temporal change. However, in the proposed formulation that follows, we prove that this is not the case and the condition of spatial continuity of the tangential electric field components across the spatial boundary of a dielectric block undergoing a stepwise temporal change of permittivity is indeed satisfied through the generation of electromagnetic pulses on the boundary, which propagate on either side of the boundary and lead to partial conversion of the electrostatic field to radiation fields, as shown schematically in Fig. 1(a).

We now study analytically the time-dependent feature introduced above. Without loss of generality and for the sake of simplicity, instead of considering a time-varying dielectric block standing in an open space like in Fig. 1, we focus on the parallel waveguide scenario illustrated in Fig. 2(a). The structure is assumed to be invariant along the $y$-direction, infinite in the $z$-direction, and bounded in the $x$-direction. The structure consists of a time-varying dielectric rectangular block of size $2L \times d$ sandwiched between two horizontal impenetrable [e.g., perfectly electric conducting (PEC)] walls. The regions outside the time-varying dielectric block are assumed to be filled with air. In order to create a static electric field distribution inside this structure, we can imagine connecting the metallic plates to a constant (DC) voltage source (i.e., battery). After the initial transient has passed, the voltage source establishes an $x$-directed uniform static electric field inside the structure given by $e_x^s = V/d$, with $V$ and $d$ the voltage across the PEC plates and the separation between them, respectively. Once $e_x^s$ has been established inside the whole structure and the system has reached a steady state, the permittivity of the dielectric block is made to vary in time rapidly. We assume that the relative permittivity ($\varepsilon_r$) rapidly switches from one value $\varepsilon_{r1}$ to a different one $\varepsilon_{r2}$ (both positive values greater than unity) at $t = 0$, as shown in the inset of Fig. 2(a).

Due to the symmetry of the structure in Fig. 2(a) with respect to the plane $z = 0$ and the distribution of $e_x^s$, we can analyze the response of the system by solving the reduced structure in Fig. 2(b). It consists of half of the structure to the right of the symmetry plane ($z = 0$) and a perfectly magnetic conducting (PMC) wall on that plane. The field distribution on the left side of the symmetry plane will be the mirror image of that on the right side. A suitable way of solving the problem under study is to formulate it as an initial value problem of Maxwell equations for transverse electromagnetic (TEM) waves propagating in the $z$-direction. As initial values, we consider the value of the electric field at $t = 0^+$, i.e., just after the change of the permittivity, in the air and dielectric regions, given by $e_x^{sa}(z,0^+) = e_x^s$ and $e_x^{sd}(z,0^+) = \frac{\varepsilon_{r1}}{\varepsilon_{r2}} e_x^s$, respectively. These fields are obtained by considering that across the instant $t = 0$, $e_x^{sa}$ remains continuous while $e_x^{sd}$ is discontinuous, modeling the disturbance caused by the permittivity change and its value at $t = 0^+$ is obtained by imposing the temporal continuity of **d**. Solution of Maxwell equations with these initial conditions can be achieved through the Laplace transform (a complete derivation of the fields is given in Supplementary Materials [16]), leading to the following expressions for the time dependent electric fields in the dielectric and air regions for $t > 0$:

$$e_x^a(z,t) = \frac{e_x^s \sqrt{\varepsilon_{r2}}}{1 - \sqrt{\varepsilon_{r2}}} \left( \frac{\varepsilon_{r1}}{\varepsilon_{r2}} - 1 \right) \left[ -\gamma u \left( t - \frac{1}{c_a}(z-L) \right) + (1-\gamma) \sum_{n=1}^{\infty} (\gamma)^n u \left( t - \frac{1}{c_a}(z-L) + 2nT \right) \right] + \\ + u(t) e_x^s \qquad z \geq L \tag{1}$$

$$e_x^d(z,t) = \frac{e_x^s}{\sqrt{\varepsilon_{r2}}+1} \left( 1 - \frac{\varepsilon_{r1}}{\varepsilon_{r2}} \right) \sum_{n=0}^{\infty} (\gamma)^n \left[ u \left( t - \frac{1}{c_d}(L-z) - 2nT \right) + u \left( t - \frac{1}{c_d}(z+L) + 2nT \right) \right] + \\ + u(t) \frac{\varepsilon_{r1}}{\varepsilon_{r2}} e_x^s \qquad 0 \leq z < L \tag{2}$$

with $c_a$ the speed of light in air, $c_d = c_a / \sqrt{\varepsilon_{r2}}$, $\gamma = \left( \sqrt{\varepsilon_{r2}} - 1 \right) / \left( \sqrt{\varepsilon_{r2}} + 1 \right)$ (reflection coefficient at section $z = L$), $u(\alpha)$ is the Heaviside unit function, which equals to 1 for $\alpha > 0$ and zero for $\alpha < 0$, and $T = L/c_d$ is the time of propagation between the air-dielectric interface and the magnetic wall. Both Eqs. (1) and (2) can be viewed as a superposition of static and dynamic electric fields. The term $u(t)$ represents the static electric field with amplitude equal to the corresponding region's initial value. The dynamic electric field consists of an infinite number of waves traveling along the $z$-axis with different amplitude and time delay $2T$ from each other, equal to the time it takes a wave to propagate through the dielectric slab. As expected, Eq. (1), describing the air region's electric field, contains plane waves of the form $u\left(t - \frac{1}{c_{air}}(z-L) + 2nT\right)$, which propagate outward from the air-dielectric interface ($z = L$). On the other hand, Eq. (2), describing the dielectric region's electric field, contains an infinite number of plane waves of the form $u\left(t - \frac{1}{c_{air}}(L \pm z) \pm 2nT\right)$, which propagate both inward and outward from the same interface, as a result of reflection on the two boundaries of the dielectric slab (the air-dielectric interface and the PMC wall). Using Eqs. (1) and (2), and time-dependent Maxwell equations, it is straightforward to also calculate the magnetic field in the dielectric and air regions. As shown in [16], in both regions, the magnetic field, as expected, is directed along the $y$-axis and is dynamic only (its expression does not containing any terms with $u(t)$). The analytical expressions for both the electric and magnetic field have been validated through comparison against numerical solutions.

To get a better understanding of the temporal evolution of this wave phenomenon and the associated energy flow, we now look at the spatial distribution of the electric field and of the instantaneous Poynting vector inside the waveguide at four different time instants ($t = 0.1T, 0.7T, 1.3T, 2.2T$). Hereafter, we assume that the length of the dielectric block is $L$ and its relative permittivity changes from $\varepsilon_{r1} = 8$ to $\varepsilon_{r2} = 4$. The spacing and the DC voltage between the waveguide metallic plates are $d = 1\text{mm}$ and $V = 1\text{V}$, respectively. With this spacing and voltage, at the steady state the electrostatic field inside the waveguide is uniform and equal to $1\text{V}/mm$. As mentioned earlier, the change of the dielectric-block permittivity occurs at $t = 0$. According to the temporal boundary conditions, this change perturbs the electrostatic field distribution in the dielectric region, transforming its value from $1\text{V}/mm$ to $2\text{V}/mm$, while the value of the electrostatic field in the air region is still equal to $1\text{V}/mm$. Now, observing the spatial distribution of the electric field a bit after $t = 0$, say $t = t_1 = 0.1T$, shown in Fig. 3(a), we notice that, there is a region across the air-dielectric interface where the electric field is between $1\text{V}/mm$ nor $2\text{V}/mm$. Quite importantly, the electric field is continuous across the spatial boundary, as required by the spatial boundary conditions. From Eqs. (1) and (2), we can identify this field with the plane waves ($2/3u(t_1-(z-L)/c_a)$ and $-1/3u(t_1-(L-z)/c_d)$) propagating in opposite directions from the dielectric-air interface, with velocities equal to the wave velocities in the two media. It is now interesting to explore the energy flow into the waveguide. Fig. 3(e) displays the $z$-component of the instantaneous Poynting vector ($S_z$) at the same time instance ($t = t_1$) as Fig. 3(a). The other components of the Poynting vector are equal to zero as we are considering TEM plane waves with respect to the z axis. By comparing Figs. 3(a) and 3(e), one can observe that, as expected, $S_z$ is different than zero only in the region between the two wavefronts and points in the positive z direction, showing power flow from the dielectric medium to air. Hence, the right-going plane wave, propagating in the air region along the positive $z$-axis, is a standard forward wave. On the other hand, the left-going plane wave, propagating in the dielectric region along the negative z axis, is a backward wave. At $t = t_2 = 0.7T > t_1$, the two plane waves ($2/3u(t_2-(z-L)/c_a)$ and $-1/3u(t_2-(L-z)/c_d)$) have propagated even further from the air-dielectric interface as highlighted by the position of their wavefronts (see Fig. 3(b)), and a greater presence of electromagnetic energy movement is observed into the waveguide (see Fig. 3(f)). At $t = T$ (T the time delay introduced by the dielectric block), the plane wave, $-1/3 u(T-(L-z)/c_d)$, propagating inside the dielectric, hits the perfect magnetic wall located at $z = 0$, generating a second plane wave, represented by $-1/3 u(t-(z+L)/c_d)$ in Eq. (2), which is added to the electric field established by the first wave $-1/3 u(t_2-(L-z)/c_d)$, resulting in a filed distribution as in Fig. 3(c). On the other hand, in the air region there is still the same plane wave ($2/3 u(t_3-(z-L)/c_a)$) that was propagating in it during the previous time interval ($0 < t < T$), which is perturbing the electrostatic field distribution established by the voltage source. From Fig. 3(g), which displays the distribution of $S_z$ into the waveguide at the same time instant ($t = t_3$) of the electric field in Fig. 3(c), we notice that $S_z$ is positive also at this time instant, showing flow of energy from the dielectric to air. At $t = 2T$, the plane wave $-1/3 u(t-(z+L)/c_d)$ propagating in the dielectric reaches the air-dielectric interface (section $z = L$) and experiences a reflection, resulting in a pulse again propagating towards the PMC wall in the dielectric region [$-2/3 u(t-(z-L)/c_a+2T)$] and another pulse that perturbs the field in the air region [$-1/9 u(t-(L-z)/c_d-2T)$], as can be seen in Fig. 3(d). At the same time, the direction of power flow is still from the dielectric to air region (see Fig. 3(h)). As time passes, due to the spatial boundaries located at $z = 0$ (perfect magnetic wall) and $z = L$ (air-dielectric interface), the number of plane waves propagating into the waveguide increases as described by the summations in Eqs. (1) and (2). However, all waves in both regions always carry power from the dielectric to air region. At $t \to \infty$, the amplitudes of the plane waves, given by $|\gamma|^n$, decay to zero, since $|\gamma| < 1$, and the electric field converges to the uniform static electric field ($e_x^s$) established inside the waveguide by the DC voltage source. This point highlights the transitory nature of the wave phenomenon induces by the change of the dielectric-block permittivity, which generates a series of plane waves whose amplitude vanishes after a transient time and the system returns to the initial steady state. The Fourier transform of the electric field in the air region (Eq. (1)) is given by

$$E_x^a(z,\omega) = \frac{2e_x^s\sqrt{\varepsilon_{r2}}}{1+\sqrt{\varepsilon_{r2}}}\left(\frac{\varepsilon_{r1}}{\varepsilon_{r2}}-1\right)\frac{\sin(\omega T)}{\omega}\frac{e^{-i\omega\left(\frac{z-L}{c_a}+T\right)}}{1-\gamma e^{-i2\omega T}} + e_x^s\frac{1}{i\omega}$$

with $\omega$ the angular frequency and $i=\sqrt{-1}$. From the latter equation, one can observe that the magnitude of $E_x^a(z,\omega)$ assumes the shape of a *sinc* function with the main lobe width related to $T = L/c_d$. Thus, the larger T, which results from a long and/or a high $\varepsilon_{r2}$ dielectric block, the narrower the main lobe width will be. More details on $E_x^a(z,\omega)$ are provided in Supplementary Materials [16].

We now turn to the discussion of energy exchanges undergoing into the system as a consequence of the permittivity change of the dielectric block. To this end, we consider $W = |\mathbf{d}|^2 L/(2\varepsilon) \left[J/m^2\right]$, which is the electric energy density stored in an electrostatic field distribution in a dispersionless dielectric medium with permittivity $\varepsilon$ and length $L$. By means of the previous equation, it is straightforward to derive the electric energy densities associated to the electrostatic field in the dielectric block before ($t = 0^-$) and after ($t = 0^+$) the permittivity change. These energy densities are $W_{0^-} = \frac{1}{2}\varepsilon_0\varepsilon_{r1}L\left(\frac{V}{d}\right)^2$ and $W_{0^+} = \frac{1}{2}\varepsilon_0\frac{\varepsilon_{r1}^2}{\varepsilon_{r2}}L\left(\frac{V}{d}\right)^2$ at $t = 0^-$ and $t = 0^+$, respectively, with the latter derived from the continuity of $\mathbf{d}$ across the temporal boundary at $t = 0$. Taking the difference between $W_{0^+}$ and $W_{0^-}$, we find $\Delta W_0 = \frac{1}{2}\varepsilon_0\varepsilon_{r1}L\left(\frac{\varepsilon_{r1}}{\varepsilon_{r2}}-1\right)\left(\frac{V}{d}\right)^2$, which gives the energy supplied to or taken from the system upon the change of the permittivity by an external agent performing this change. For $\varepsilon_{r1} > \varepsilon_{r2}$, which is the case discussed so far with $\varepsilon_{r1} = 8$ and $\varepsilon_{r2} = 4$, $\Delta W_0 > 0$ implying that the permittivity change increases the electric energy density inside the dielectric block. This means that the external agent provides energy to the dielectric block when changing its permittivity. The second mechanism of energy exchange occurs between the electrostatic energy in the dielectric block and the electromagnetic energy propagating into the system. The energy balance of this mechanism can be expressed as the difference between the electric energy density in the dielectric block at $t = 0^+$ and after the transient time, that is at $t \to \infty$. As discussed earlier, at $t \to \infty$ the system has returned to a steady state with a uniform static field distribution ($V/d$) and, as a result, the electric energy density in the dielectric block is given by $W_\infty = \frac{1}{2}\varepsilon_0\varepsilon_{r2}L\left(\frac{V}{d}\right)^2$. Therefore, the energy balance is $\Delta W_\infty = \frac{1}{2}\frac{\varepsilon_0}{\varepsilon_{r2}}L\left(\varepsilon_{r2}^2 - \varepsilon_{r1}^2\right)\left(\frac{V}{d}\right)^2$. For $\varepsilon_{r1} > \varepsilon_{r2}$, $\Delta W_\infty < 0$ meaning that the dielectric block loses part of its energy, which is released in the form of electromagnetic energy traveling away from the dielectric block. Indeed, as discussed above and shown in Figs. 3(e), 3(f), 3(g) and 3(h), the plane waves, produced by the permittivity change, propagating into the waveguide have the Poynting vector along the positive $z$-axis and, therefore, they carry the energy from the dielectric region to the air region. The total radiated energy density, given by the infinite set of plane waves, is equal to $-\Delta W_\infty$ [16], highlighting that the difference between the electrostatic energy in the dielectric block at $t = 0^+$ and $t \to \infty$ is fully released in the form of electromagnetic energy in the air region.

So far, the discussion has focused on the case $\varepsilon_{r1} > \varepsilon_{r2}$. A complete and detailed discussion of the opposite case, $\varepsilon_{r1} < \varepsilon_{r2}$, is provided in Supplementary Material [16]. However, it is worth highlighting the differences between the two cases, mostly from the energy flow standpoint. By increasing the permittivity of the dielectric block, the electric energy in it decreases. In fact, in contrast to the case with $\varepsilon_{r1} > \varepsilon_{r2}$, the energy balance of the first mechanism of the energy exchange is negative ($\Delta W_0 < 0$), i.e., the external agent responsive for the change of the permittivity receives energy from the system. Opposite behavior, with respect to the case with $\varepsilon_{r1} > \varepsilon_{r2}$, is also observed regarding the second mechanism of the energy exchange. With $\varepsilon_{r1} < \varepsilon_{r2}$, $\Delta W_\infty > 0$ meaning that the dielectric block gains electric energy. This energy is provided by the plane waves produced by the permittivity change while propagating into the waveguide. In contrast to the case with $\varepsilon_{r1} > \varepsilon_{r2}$, these plane waves have the Poynting vector directed along the negative z-axis [16], and, as a result, they carry energy from the air to the dielectric region. The air region is temporally depleted from its energy by backward waves propagating away the dielectric region. This energy is transferred to the dielectric region and is restored to its original value at $t \to \infty$. More details about this case are provided in Supplementary Materials [16].

In the last part of this letter, we will focus on the spatial-continuity condition of the tangential components of electric and magnetic fields across the air-dielectric interface in the waveguide in Fig. 2(a) after the change of the permittivity of the dielectric block ($t > 0$) from $\varepsilon_{r1} = 8$ to $\varepsilon_{r2} = 4$. Figs. 3(a), 3(b), 3(c) and 3(d) clearly show that the electric field is continuous across the air-dielectric interface (section $z = L$) at four different time instants. Now, we aim to show that the electric field ($e_x$) and the magnetic field ($h_y$) are continuous across $z = L$ at any time instant. To this end, we plot $e_x$ (from Eqs. (1) and (2)) and $h_y$ (from Eqs. in [16]) as a function of time at different observation points before and after the section $z = L$. Figs. 4(a) and 4(b) show the temporal evolution of $e_x$ and $h_y$, respectively, evaluated at the locations $z = 0.9L$ and $z = 1.1L$. As discussed before, the amplitude of $e_x$ decays as time passes and after the transient time its value returns to be one. Analogous behavior is observed for $h_y$ [Fig. 4(b)]; its amplitude converges to zero after the transient time. As can be seen from Fig. 4(a) (Fig. 4(b)), the temporal profile of $e_x$ ($h_y$) evaluated at $z = 0.9L$ and $z = 1.1L$ do not overlap with each other. Therefore, $e_x$ and $h_y$ seem to be discontinuous across the air-dielectric interface ($z = L$). However, moving the observation points even closer to $z = L$, such as $z = 0.99999L$ and $z = 1.00001L$, the temporal profiles of $e_x$ ($h_y$) at these two locations overlap with each other as shown in Fig. 4(c) (Fig. 4(d)). More rigorously, it can be mathematically shown [16] that for time instants very close to $t = 0$, yet with finite values, there is always a region around the air-dielectric interface within which the fields are continuous, as they should be. These results show that that the condition of the spatial continuity of the tangential of the fields across a time-varying dielectric block embedded in a uniform static electric field is expectedly satisfied.

In conclusion, we have shown, using an analytical approach, how the static-to-dynamic field conversion can be achieved through a time-varying dielectric block embedded in an electrostatic field distribution. Such dielectric block experiencing a change of its permittivity in time gives rise to radiation fields. These fields start propagating into the system from the spatial boundary of the time-varying material enforced by the continuity of the electric and magnetic fields' tangential components at such boundary. The radiation fields carry the energy away (into) the time-varying dielectric block by decreasing (increasing) its permittivity. The results presented here might find applications in the design of electromagnetic sources in integrated systems, similar to sources that have been demonstrated in the past in plasmas.

**Acknowledgment**:

N.E. and M.F. acknowledge partial support from the Simons Foundation/Collaboration on Symmetry-Driven Extreme Wave Phenomena.

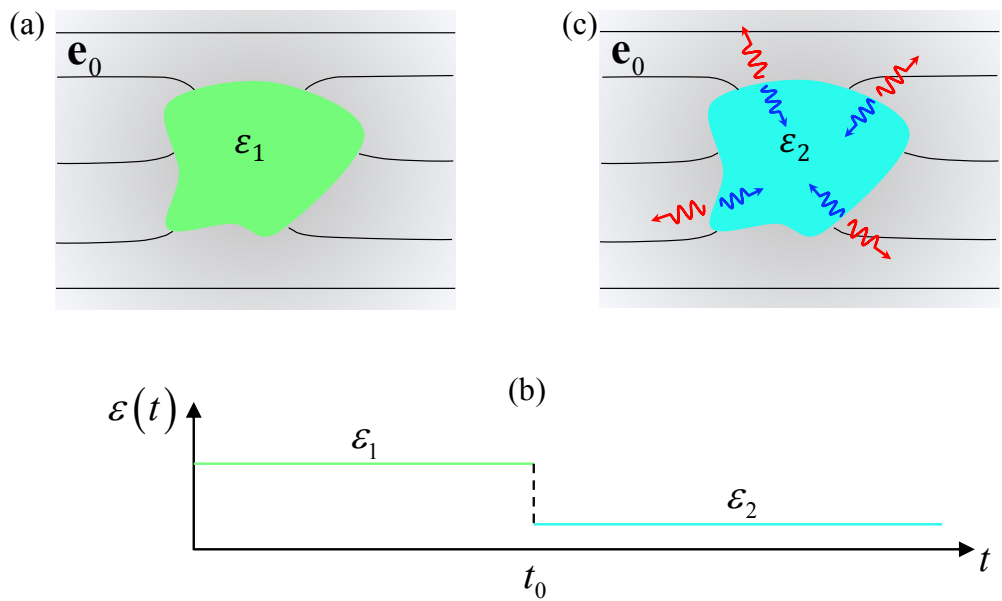

**Figure 1.** Conceptual representation of static-to-radiative field conversion based on time-varying media. (a) A time-varying dielectric block embedded in a uniform static electrostatic field $\mathbf{e}_0$. (b) Temporal profile of permittivity, which changes from a positive value $\varepsilon_1$ to another positive value $\varepsilon_2$ at $t = t_0$. (c) Sketch of the wave phenomenon generates by the change of the permittivity of the dielectric block.

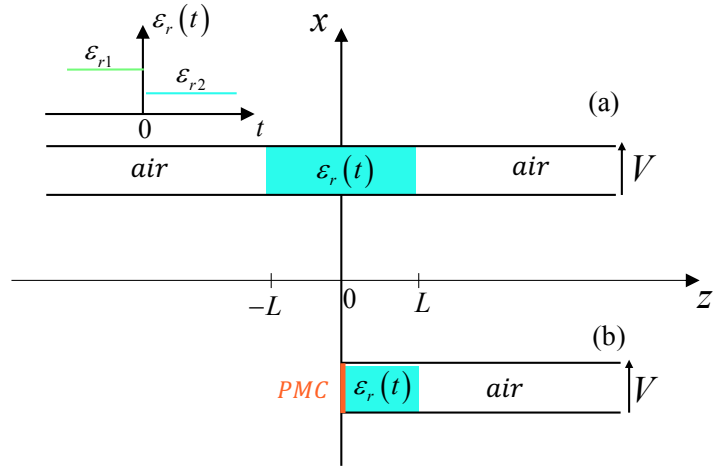

**Figure 2. Parallel waveguide** scenario under study. (a) A time-varying dielectric rectangular block of size $2L \times d$ is placed at center of a waveguide filled with air. The PEC walls of the waveguide are connected to a DC voltage source establishing uniform static electric field inside the waveguiding structure. The time-dependent relative permittivity ($\varepsilon_r(t)$) of the dielectric is shown in the inset. (b) The waveguide is terminated with a PMC wall.

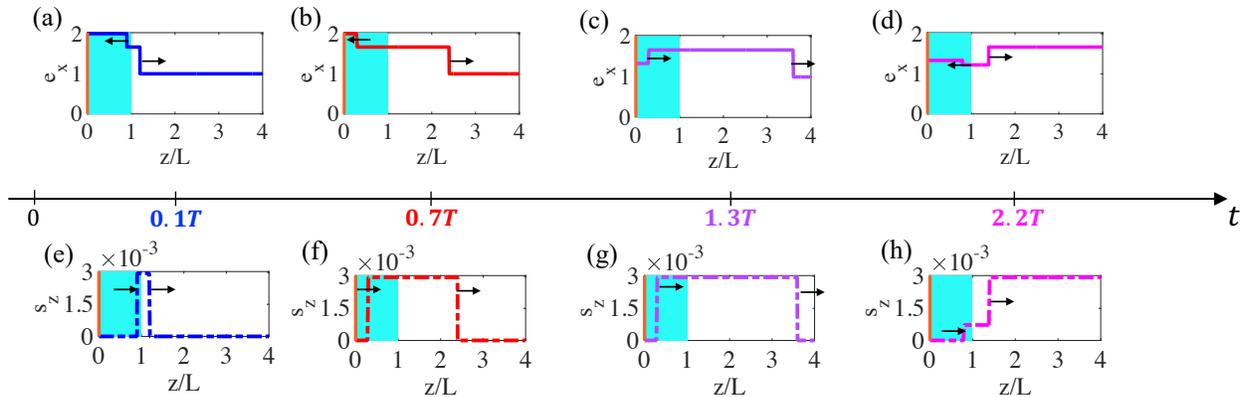

**Figure 3.** Spatial distribution of the electric field, (a), (b), (c), and (d), and the Poynting vector, (e), (f), (g), and (h), at time instants $t = 0.1T$, $0.7T$, $1.3T$, and $2.2T$ inside the waveguide of Fig. 2(b). The arrows in the top panels ((a), (b), (c), and (d)) and in the bottom panels ((e), (f), (g), and (h)) indicate the direction of propagation of the plane waves presenting into the waveguide at the considered time instants and of the corresponding Poynting vectors, respectively.

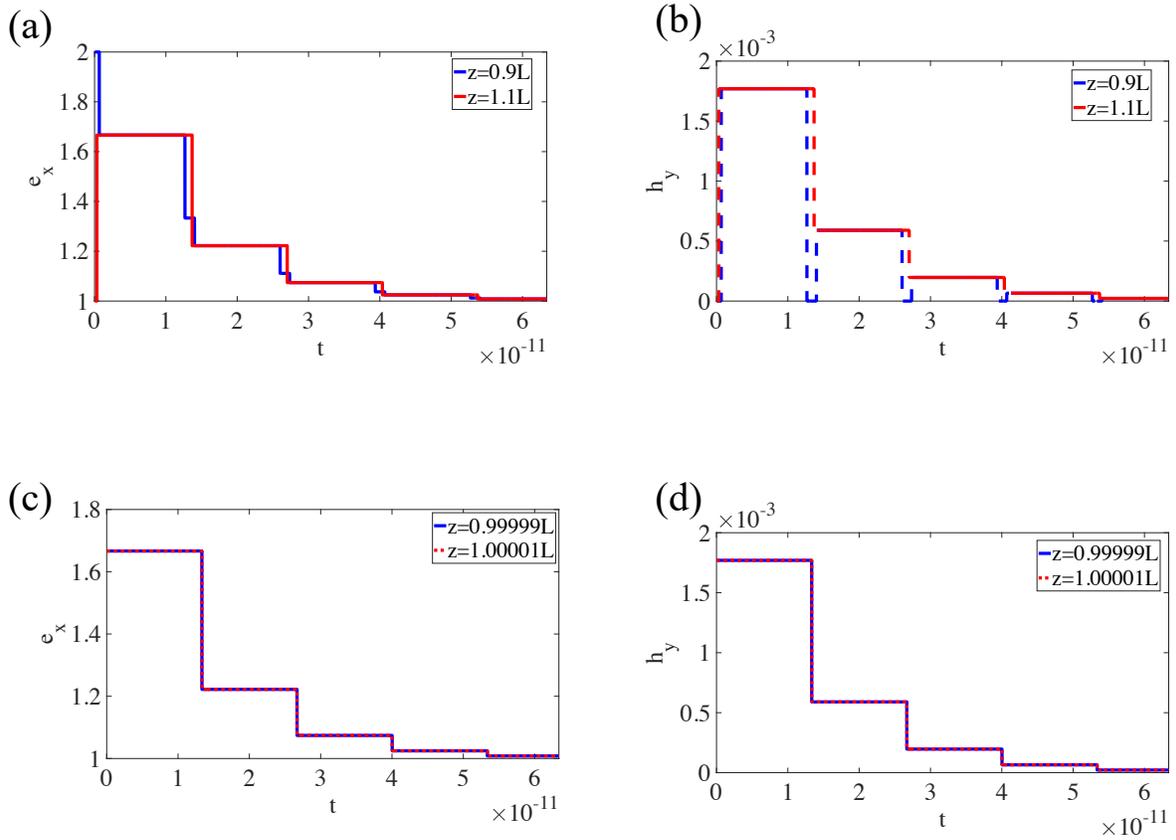

**Figure 4.** Temporal evolution of the electric field, (a) and (c), and the magnetic field, (b) and (d), at the observation points $z = 0.9L$, $z = 1.1L$ and $z = 0.99999L$, $z = 1.0001L$, respectively.

# Supplementary material for "Static-to-dynamic field conversion with time-varying media"

By Mario Junior Mencagli, Dimitrios L. Sounas, Mathias Fink, and Nader Engheta*

*Corresponding author: engheta@seas.upenn.edu

1. Introduction
2. Initial value problem for plane waves
3. Analytical solution
   3.1. Laplace domain
   3.2. Time domain
   3.3. Analytical vs numerical solutions
4. The limit $t \to \infty$
5. Continuity of electric and magnetic fields at the interface of the time-varying dielectric block
6. Increasing the permittivity of the time-varying block
7. Total radiated energy density
8. Fourier spectrum of the electric field in the air region

*1. Introduction*

We aim to provide a detailed analysis of the waveguiding system displayed in Fig. 2(b) of the main text, represented in Fig. S1, for completeness and easy access. First, we discuss the initial value problem for plane waves, which enables deriving its electric and magnetic fields given the problem's initial condition. Then, we apply the solution of such an initial value problem to solve the waveguiding system in Fig. S1. The derived analytical electric and magnetic fields expressions are obtained in the Laplace domain and then converted in the time domain through an inverse transformation. To validate the proposed formulation's accuracy, we compare the analytical solution against a numerical solution. The time development of the system for $t \to \infty$ and the spatial boundary condition at the air-dielectric interface ($z = L$) are investigated. Then, we examine the case in which the relative permittivity of the dielectric material increases. Finally, we derive and discuss the total radiated energy and the Fourier spectrum of the electric field in the air region.

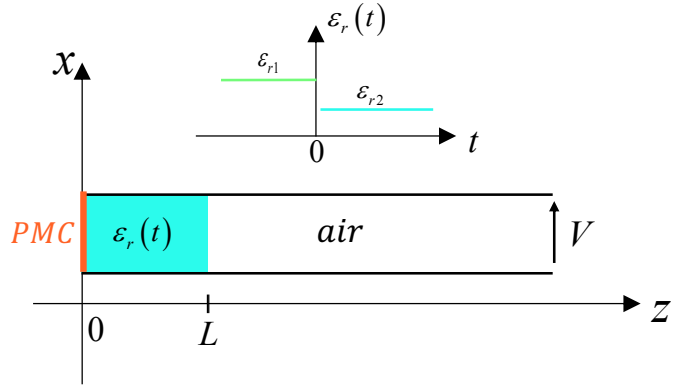

**Figure S1.** A waveguide, consisting of two horizontal PEC walls located at a distance $d$ apart, is terminated with a PMC wall at one end and a voltage source at the other end. The waveguide is filled with a time-varying dielectric and air in regions $0 \leq z \leq L$ and $z > L$, respectively. The inset shows the temporal variation of the dielectric block. The waveguiding system is invariant along the $y$-axis.

## 2. Initial value problem for plane waves

This section shows an analytical solution to the Maxwell equations for TEM waves constrained by an initial value defining the electric field's value at the initial time instant, say $t = 0^+$. Let $e_x(t,z)$ and $h_y(t,z)$ be the electric and magnetic field of a $z$-propagating TEM wave. We assume that $e_x(t,z)$ and $h_y(t,z)$ vanish for $t < 0$, and they are obviously coupled one another by the Maxwell equations

$$\frac{\partial e_x(t,z)}{\partial z} + \mu \frac{\partial h_y(t,z)}{\partial t} = 0 \tag{S1}$$

$$\frac{\partial h_y(t,z)}{\partial z} + \varepsilon \frac{\partial e_x(t,z)}{\partial t} = 0 \tag{S2}$$

with $\varepsilon$ and $\mu$ the permittivity and the permeability of the medium, respectively. The aim now is to solve these coupled differential equations for $t > 0$ by taking into account the electric field's initial value at $t = 0^+$. A suitable way to introduce this initial value is to Laplace transform $e_x(t,z)$ and $h_y(t,z)$ with respect to $t$: $E_x(s,z) = \int_0^\infty e_x(t,z) e^{-st} dt$, $H_y(s,z) = \int_0^\infty h_y(t,z) e^{-st} dt$. Applying these transformations to Eqs. (S1) and (S2), and using the differentiation properties of the Laplace transform [1] we get

$$\frac{\partial E_x(s,z)}{\partial z} + s\mu H_y(s,z) = 0 \tag{S3}$$

$$\frac{\partial H_y(s,z)}{\partial z} + \varepsilon s E_x(s,z) - \varepsilon e_x(0^+,z) = 0 \tag{S4}$$

with $e_x(0^+,z)$ the electric field's value at $t=0^+$. Eliminating $H_y(s,z)$ from the previous equations, an ordinary, inhomogeneous, differential equation for $E_x(s,z)$ is obtained, in which $z$ is treated as a parameter,

$$\frac{\partial^2 E_x(z,s)}{\partial z^2} - k^2 E_x(z,s) = \xi(z,s) \tag{S5}$$

with $k = s/c$, $c = 1/\sqrt{\varepsilon\mu}$, and $\xi(s,z) = -\frac{k^2}{s} e_x(0^+,z)$. Eq. (S5) is satisfied by [1]

$$E_x(s,z) = c_1 e^{kz} + c_2 e^{-kz} + \frac{e^{kz}}{2k}\int e^{-kz}\xi(z,s)dz - \frac{e^{-kz}}{2k}\int e^{kz}\xi(z,s)dz \tag{S6}$$

,with $c_1$ and $c_2$ constants that are to be determined to satisfy the spatial boundary conditions, which represents the electric field of a plane wave in the Laplace domain constrained by the initial value $e_x(z,0^+)$.

## 3. Analytical solution

Here, building on the previous section's discussion, we provide an analytical solution of the transient electric and magnetic fields resulting from the abrupt temporal change of the relative permittivity ($\varepsilon_r(t)$) of the dielectric block in the waveguiding system of Fig. S1. As discussed in the main text, it is assumed that $\varepsilon_r(t)$ suddenly varies between two positive values at $t=0$, once the DC voltage source, connected to the metallic plates, has already established a uniform static electric field ($e_x^s$) inside the waveguide. The change of permittivity results in a discontinuity of electrostatic field distribution inside the dielectric block, which can be derived by applying the continuity of **d** at $t = 0$, as $\mathbf{d}_{-\delta} = \mathbf{d}_{\delta}$ in the limit when $\delta \to 0$. Based on this, the electrostatic field inside the dielectric block at $t = 0^+$ (right after the permittivity change) reads

$$e_x^{sd} = \frac{\varepsilon_{r1}}{\varepsilon_{r2}} e_x^s \tag{S7}$$

On the other hand, the permittivity in the air region does not change, and the electrostatic field stays continuous across $t = 0$ and its value at $t = 0^+$ reads

$$e_x^{sa} = e_x^s \tag{S8}$$

These initial values ($e_x^{sd}$ and $e_x^{sa}$), which model the disturbance caused by the permittivity change of a portion of the medium inside the waveguide, can be used in the initial value problem discussed in the previous section to study the temporal evolution of the system for $t > 0$.

### 3.1. Laplace domain

Plugging the initial values of Eqs. (S7) and (S8) into Eq. (S6), we can readily derive the electric field in the dielectric and air regions in the Laplace domain

$$E_x^d(s,z) = A e^{k_d z} + B e^{-k_d z} + \frac{1}{s} \frac{\varepsilon_{r1}}{\varepsilon_{r2}} e_x^s \quad 0 \leq z \leq L \tag{S9}$$

$$E_x^a(s,z) = C e^{k_a z} + D e^{-k_a z} + \frac{1}{s} e_x^s \quad z \geq L \tag{S10}$$

with $c_a$ the speed of light in free space, $k_{a,d} = s/c_{a,d}$, and $c_d = c_a/\sqrt{\varepsilon_{r2}}$. The magnetic field in both regions ($H_y^d$ and $H_y^a$) can be derived from Eq. (S3). The four unknown constants ($A, B, C, D$) can be obtained by applying the conservation of the transverse components of the electric and magnetic fields at the interface $z = L$ (see Fig. S1), zeroing the tangential components of the magnetic field at the interface $z = 0$, and the radiation condition at $z \to \infty$. In doing so, we obtain the following expressions for the electric and magnetic fields in the two regions:

Electric field:

$$E_x^d(s,z) = e_x^s \frac{1}{s}\left(1 - \frac{\varepsilon_{r1}}{\varepsilon_{r2}}\right)\left(\frac{k_a\left(e^{k_d z} + e^{-k_d z}\right)}{k_a\left(e^{k_d L} + e^{-k_d L}\right) + k_d\left(e^{k_d L} - e^{-k_d L}\right)}\right) + \frac{1}{s}\frac{\varepsilon_{r1}}{\varepsilon_{r2}} e_x^s \quad 0 \leq z \leq L \quad (S11)$$

$$E_x^a(s,z) = e_x^s \frac{e^{k_a(L-z)}}{s}\left(-1 + \frac{\varepsilon_{r1}}{\varepsilon_{r2}}\right)\left(\frac{k_d\left(e^{k_d L} - e^{-k_d L}\right)}{k_a\left(e^{k_d L} + e^{-k_d L}\right) + k_d\left(e^{k_d L} - e^{-k_d L}\right)}\right) + \frac{1}{s} e_x^s \quad z > L \quad (S12)$$

Magnetic field:

$$H_y^d(s,z) = e_x^s \frac{k_a}{\mu_0 s^2}\left(-1 + \frac{\varepsilon_{r1}}{\varepsilon_{r2}}\right)\left(\frac{k_d\left(e^{k_d z} - e^{-k_d z}\right)}{k_a\left(e^{k_d L} + e^{-k_d L}\right) + k_d\left(e^{k_d L} - e^{-k_d L}\right)}\right) \quad 0 \leq z \leq L \quad (S13)$$

$$H_y^a(s,z) = e_x^s \frac{e^{k_a(L-z)}}{\mu_0 s^2}\left(-1 + \frac{\varepsilon_{r1}}{\varepsilon_{r2}}\right)\left(\frac{k_a k_d\left(e^{k_d L} - e^{-k_d L}\right)}{k_a\left(e^{k_d L} + e^{-k_d L}\right) + k_d\left(e^{k_d L} - e^{-k_d L}\right)}\right) \quad z > L \quad (S14)$$

with $\mu_0$ the free space permeability. The remaining difficulty is transforming the expressions of the derived fields in the Laplace domain to the time domain, which is the subject of the next section.

*3.2. Time domain*

The most common technique to performing the inverse Laplace transform of complex functions consists of evaluating a contour integral with the residue theorem [2]. Although such technique can be applied to Eqs. (S11)-(S14), we have followed a different way: reducing such equations into a composition of functions whose inverse Laplace transform is known. Let us begin with the electric field in the dielectric region (Eq. (S11)). The right-hand side of Eq. (S11) exhibits a singular behavior in the complex $s$-plane at $s=0$ and

$s_m = \ln\gamma + i(2m+1)\pi$ with $m = 0, \pm 1, \pm 2, \ldots$, $i = \sqrt{-1}$, and $\gamma = \dfrac{\sqrt{\varepsilon_{r2}}-1}{\sqrt{\varepsilon_{r2}}+1}$ (the reflection coefficient at air-dielectric interface). It is straightforward to show that these singularities are poles of $E_x^d$ and, as $0 \leq |\gamma| < 1$, they lie in the open left half of the complex $s$-plane including the imaginary axis. Therefore, expressing $s = \sigma + i\omega$ ($\sigma$ and $\omega$ two real numbers), the existence region of $E_x^d$, which defines the set of $s$ values where the integral in the definition of Laplace transform converges [2], is given by $\sigma > 0$. Since this region does not include the imaginary axis of the $s$-plane ($\sigma = 0$), one can state that the inverse Laplace transform of $E_x^d$, corresponding to the time-domain electric field ($e_x^d(t,z)$), is only marginally stable. Nevertheless, $e_x^d(t,z)$ asymptotically (for $t \to \infty$) converges to a finite value, as will be shown later.

After discussing the existence region and the stability of $E_x^d$, we proceed to carry out its inverse Laplace transform. Substituting $k_{a,d} = s/c_{a,d}$ in Eq. (S11) and simplifying, we obtain

$$E_x^d(s,z) = \frac{e_x^s}{s}\left(1-\frac{\varepsilon_{r1}}{\varepsilon_{r2}}\right)\left(e^{\frac{s}{c_d}(z-L)} + e^{-\frac{s}{c_d}(z+L)}\right)\frac{e^{2\frac{s}{c_d}L}}{e^{2\frac{s}{c_d}L}\left(\sqrt{\varepsilon_{r2}}+1\right)+1-\sqrt{\varepsilon_{r2}}} + \frac{1}{s}\frac{\varepsilon_{r1}}{\varepsilon_{r2}}e_x^s \qquad (S15)$$

By inspecting the previous equation, with

$$Q \equiv \frac{e^{2\frac{s}{c_d}L}}{e^{2\frac{s}{c_d}L}\left(\sqrt{\varepsilon_{r2}}+1\right)+1-\sqrt{\varepsilon_{r2}}} \qquad (S16)$$

one can notice that except $Q$ the other terms are in the form for which the inverse Laplace transform is known. Replacing $e^{2\frac{s}{c_d}L}$ with $1/w$ in Eq. (S16) and rearranging, we get

$$Q = \frac{1}{1+\sqrt{\varepsilon_{r2}}}\frac{1}{1-\gamma w} \qquad (S17)$$

The second term in the previous equation is the sum of the geometric series of argument $(-\gamma w)$. Therefore, $Q$ can be rewritten as

$$Q = \frac{1}{1+\sqrt{\varepsilon_{r2}}}\left(1+\gamma w + \gamma^2 w^2 + \gamma^3 w^3 + \ldots\right) = \frac{1}{1+\sqrt{\varepsilon_{r2}}}\sum_{n=0}^{\infty}(\gamma w)^n \qquad (S18)$$

As is well-known, the geometric series converges when the absolute value of its argument is smaller than one. This condition, for the series in Eq. (S18), implies $|\gamma w| < 1$. Substituting $w = e^{-2\frac{s}{c_d}L}$, $s = \sigma + i\omega$, and $\gamma = \frac{\sqrt{\varepsilon_{r2}}-1}{\sqrt{\varepsilon_{r2}}+1}$ in the previous inequality and solving for $\sigma$, we obtain

$$\sigma > -\frac{c_d}{2L}\ln\left(\frac{\sqrt{\varepsilon_{r2}}+1}{\sqrt{\varepsilon_{r2}}-1}\right) \qquad (S19)$$

which defines the region of convergence in the $s$-plane for the series in Eq. (S18). For $\varepsilon_{r2} \geq 1$, which is the case considered in this work, the right-hand side of the previous inequality will always be a negative real number. As a result, the inequality in (S19) holds for any complex value inside the existence region of $E_x^d$, which, as discussed above, is given by $\sigma > 0$, and $Q$ in Eq. (S15) can be expressed as in Eq. (S18) to perform the inverse Laplace transform of $E_x^d$ (Eq. (S15)). Substituting Eq. (S18) in Eq. (S15) and rearranging, we obtain

$$E_x^d(s,z) = \frac{e_x^s}{\sqrt{\varepsilon_{r2}}+1}\left(1-\frac{\varepsilon_{r1}}{\varepsilon_{r2}}\right)\frac{1}{s}\left[\sum_{n=0}^{\infty}\gamma^n\left(e^{\frac{s}{c_d}(z-(2n+1)L)}+e^{-\frac{s}{c_d}(z+(2n+1)L)}\right)\right] + \frac{1}{s}\frac{\varepsilon_{r1}}{\varepsilon_{r2}}e_x^s \quad (S20)$$

Now, the time domain expression of $E_x^d$ can be easily carried out through the standard Laplace transform table. In doing so, we obtain

$$e_x^d(t,z) = \frac{e_x^s}{\sqrt{\varepsilon_{r2}}+1}\left(1-\frac{\varepsilon_{r1}}{\varepsilon_{r2}}\right)\sum_{n=0}^{\infty}\gamma^n\left[u\left(t-\frac{1}{c_d}(L-z)-2nT\right)+u\left(t-\frac{1}{c_d}(z+L)-2nT\right)\right] +$$
$$+ u(t)\frac{\varepsilon_{r1}}{\varepsilon_{r2}}e_x^s \qquad 0 \leq z < L \quad (S21)$$

Following the procedure discussed to derive $e_x^d(z,t)$, we have carried out the inverse Laplace transform of the electric field in the air region (Eq. (S12)) and its time-domain expression reads

$$e_x^a(t,z) = \frac{e_x^s\sqrt{\varepsilon_{r2}}}{1-\sqrt{\varepsilon_{r2}}}\left(\frac{\varepsilon_{r1}}{\varepsilon_{r2}}-1\right)\left[-\gamma u\left(t-\frac{1}{c_a}(z-L)\right)+(1-\gamma)\sum_{n=1}^{\infty}\gamma^n u\left(t-\frac{1}{c_a}(z-L)-2nT\right)\right] +$$
$$+ u(t)e_x^s \qquad z \geq L \quad (S22)$$

Regarding the inverse Laplace transform of the magnetic field in the air and dielectric regions (Eqs. (S13) and (S14)), the procedure adopted for the electric field can be followed. However, as we have already derived the time-domain electric field in both regions, it is more convenient to go through the time-dependent Maxwell equations. Using Eqs. (S21) and (S22) with the Maxwell equation (S1), one can derive the time-dependent magnetic field in both regions

$$h_y^d(t,z) = \frac{1}{\mu_0 c_d} \frac{e_x^s}{\sqrt{\varepsilon_{r2}}+1}\left(1-\frac{\varepsilon_{r1}}{\varepsilon_{r2}}\right)\sum_{n=0}^{\infty}\gamma^n\left[-u\left(t-\frac{L-z}{c_d}-2nT\right)+u\left(t-\frac{z+L}{c_d}-2nT\right)\right] \quad 0 \le z < L \quad (S23)$$

$$h_y^a(t,z) = \frac{e_x^s}{\mu_0 c_a}\frac{\sqrt{\varepsilon_{r2}}}{1-\sqrt{\varepsilon_{r2}}}\left(\frac{\varepsilon_{r1}}{\varepsilon_{r2}}-1\right)\left[-\gamma u\left(t-\frac{z-L}{c_a}\right)+(1-\gamma)\sum_{n=1}^{\infty}\gamma^n u\left(t-\frac{z-L}{c_a}-2nT\right)\right] \quad z \ge L \quad (S24)$$

with $\mu_0$ the free space permittivity.

It is important to note that when the dielectric block is changed to free space ($\varepsilon_{r2}=1$), the spatial boundary at the section $z=L$ disappears, and as a result, the reflection coefficient ($\gamma$) at the same section becomes zero. With this setup, Eqs. (S21), (S22), (S23), and (S24) reduce to

$$e_x^d(t,z) = \frac{e_x^s}{2}(1-\varepsilon_{r1})\left[u\left(t-\frac{L-z}{c_a}\right)+u\left(t-\frac{z+L}{c_a}\right)\right]+\varepsilon_{r1}e_x^s u(t) \quad 0 \le z < L \quad (S25)$$

$$e_x^a(t,z) = \frac{e_x^s}{2}(\varepsilon_{r1}-1)\left[u\left(t-\frac{z-L}{c_a}\right)-u\left(t-\frac{z+L}{c_a}\right)\right]+e_x^s u(t) \quad z \ge L \quad (S26)$$

$$h_y^d(t,z) = \frac{e_x^s}{2\mu_0 c_a}(1-\varepsilon_{r1})\left[-u\left(t-\frac{L-z}{c_a}\right)+u\left(t-\frac{z+L}{c_a}\right)\right] \quad 0 \le z < L \quad (S27)$$

$$h_y^a(t,z) = \frac{e_x^s}{2\mu_0 c_a}(\varepsilon_{r1}-1)\left[u\left(t-\frac{z-L}{c_a}\right)-u\left(t-\frac{z+L}{c_a}\right)\right] \quad z \ge L \quad (S28)$$

respectively. Comparing Eqs. (S25)-(S28) with Eqs. (S21)-(S24), one can observe that, as expected, the multiple reflection process, undergoing for $\varepsilon_{r2} > 1$ (see discussion in the main text), no longer happen. For $0 < t < L/c_a$, the dynamic electric field into the waveguiding system can be identified is described with the plane waves:

$\frac{e_x^s}{2}(1-\varepsilon_{r1})u\left(t-\frac{L-z}{c_a}\right)$ and $\frac{e_x^s}{2}(\varepsilon_{r1}-1)u\left(t-\frac{z-L}{c_a}\right)$ (see Eqs. (S25) and (S26)), propagating in opposite direction from section $z = L$. The former wave, hitting the PMC wall at $t = L/c_a$, gives rise to another wave, $\frac{e_x^s}{2}(\varepsilon_{r1}-1)u\left(t-\frac{z+L}{c_a}\right)$ (see Eq. (S25)), propagating in the same region ($0 \leq z \leq L$) but in the opposite direction (along the positive $z$-axis). This wave reaches the section $z = L$ at $t = 2L/c_a$, and as $\gamma = 0$, in contrast to the case with $\varepsilon_{r2} > 1$, it does not experience any reflections (no reflected waves) and results in the wave $\frac{e_x^s}{2}(\varepsilon_{r1}-1)u\left(t-\frac{z+L}{c_a}\right)$ (see Eq. (S26)) propagating in the region $z \geq L$ for $t > 2L/c_a$.

### *3.3. Analytical vs numerical solutions*

A numerical simulation of the waveguiding system in Fig. S1 has been carried out using the time domain solver of the commercial software COMSOL Multiphysics®. A lumped port, connected on the opposite side of the PMC wall, has been used to establish an electrostatic field equal to $e_x^s = 1\text{V}/\text{mm}$ between the PEC walls at distance $d = 1\text{mm}$ apart. The relative permittivity of the dielectric block, which is L=3mm long, changes abruptly from $\varepsilon_{r1} = 8$ to $\varepsilon_{r2} = 4$. This sudden change has been modeled through a step-like function with two continuous derivatives to ensure the convergence of the simulation. Figs. S2(a) and S2(b) compare the analytical and numerical distribution of the electric and magnetic fields as a function of time at locations $z = 1.5$mm and $z = 20$mm, respectively. The numerical solutions (continuous green lines) agree well with the analytical solutions (continuous blue lines) obtained from Eqs. (S21)-(S24). The numerical solutions are slightly oscillating compared to the analytical solutions. This small difference between the two solutions can be attributed to an intrinsic less accuracy in the numerical solutions.

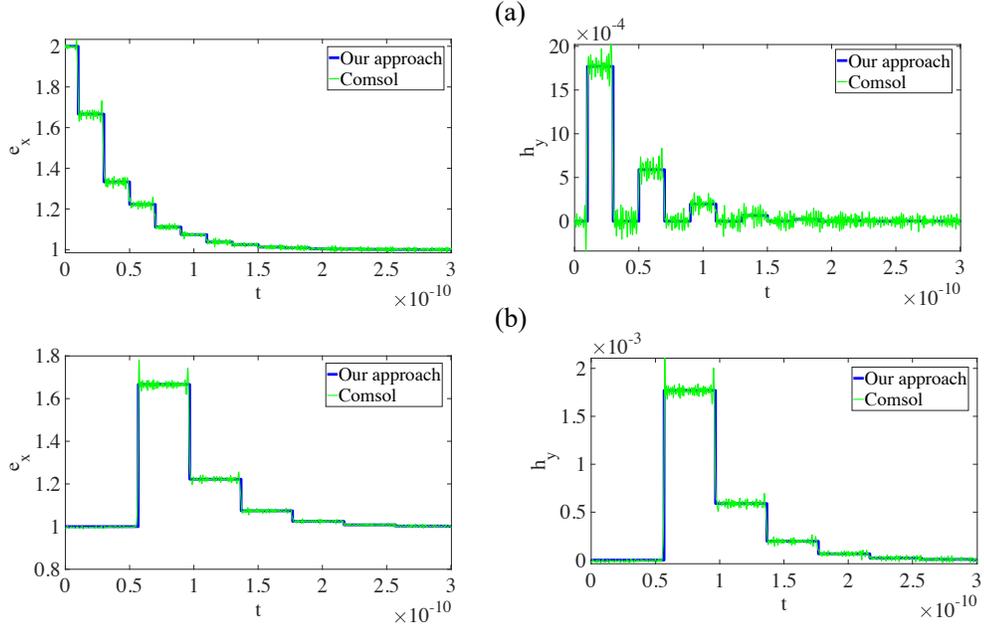

**Figure S2.** Analytical and numerical distribution of the electric and magnetic fields at the location $z = 1.5$mm (a) and $z = 20$mm (b).

## 4. *The limit* $t \to \infty$

In this section, we discuss the time evolution of the fields for $t \to \infty$. Recalling that $\lim_{t \to \infty} u(t-a) = 1$ with $a \in \mathbb{R}$, Eqs. (S21) and (S22), for $t \to \infty$, reduce to

$$e_x^d(t \to \infty, z) = \frac{2e_x^s}{\sqrt{\varepsilon_{r2}}+1}\left(1-\frac{\varepsilon_{r1}}{\varepsilon_{r2}}\right)\sum_{n=0}^{\infty}\gamma^n + \frac{\varepsilon_{r1}}{\varepsilon_{r2}}e_x^s \quad 0 \leq z < L \tag{S29}$$

S12

$$e_x^a(t \to \infty, z) = e_x^s \frac{\sqrt{\varepsilon_{r2}}}{1-\sqrt{\varepsilon_{r2}}} \left( \frac{\varepsilon_{r1}}{\varepsilon_{r2}} - 1 \right) \left[ -\gamma + (1-\gamma)\sum_{n=0}^{\infty}\gamma^n - (1-\gamma) \right] + e_x^s \quad z \geq L \tag{S30}$$

Substituting $\sum_{n=0}^{\infty}\gamma^n = \frac{1}{1-\gamma}$ (geometric series sum) and $\gamma = \frac{\sqrt{\varepsilon_{r2}}-1}{\sqrt{\varepsilon_{r2}}+1}$ (reflection coefficient at $z = L$) in the previous equations and simplifying, we obtain $e_x^d(t \to \infty, z) = e_x^a(t \to \infty, z) = e_x^s$, showing that the electric field distribution into the waveguide at $t \to \infty$ converges to a finite value, coincident with the initial electrostatic field value ($e_x^s$). Following a similar procedure for the magnetic field, it is straightforward to show that $h_y^d(t,z)$ and $h_y^a(t,z)$ in Eqs. (S23) and (S24) converge to zero for $t \to \infty$. Thus, we conclude that after the transient fields induced by the change of the dielectric block permittivity have passed, the system returns to its initial steady state.

5. ***Continuity of electric and magnetic fields at the interface of a time-varying dielectric block***

   Here, we examine the continuity of the total tangential components of the electric and magnetic fields across the interface of the time varying dielectric block after the change of its permittivity ($t > 0$). First, let us evaluate the derived expressions of the electric and magnetic fields in the dielectric and air regions at the location of the interface between the two regions ($z = L$). Plugging $z = L$ in Eqs. (S21) and (S22), we get:

$$e_x^d(t,L) = \frac{e_x^s}{\sqrt{\varepsilon_{r2}}+1}\left(1 - \frac{\varepsilon_{r1}}{\varepsilon_{r2}}\right)\sum_{n=0}^{\infty}\gamma^n\left[u(t-2nT)+u(t-2T(1+n))\right] + \frac{\varepsilon_{r1}}{\varepsilon_{r2}}e_x^s u(t) \tag{S31}$$

$$e_x^a(t,L) = e_x^s \frac{\sqrt{\varepsilon_{r2}}}{1-\sqrt{\varepsilon_{r2}}}\left(\frac{\varepsilon_{r1}}{\varepsilon_{r2}} - 1\right)\left[-\gamma u(t) + (1-\gamma)\sum_{n=1}^{\infty}\gamma^n u(t-2nT)\right] + e_x^s u(t) \tag{S32}$$

Subtracting Eq. (S32) from Eq. (S31) and rearranging, we obtain

$$e_x^d(t,L) - e_x^a(t,L) = \frac{e_x^s}{\sqrt{\varepsilon_{r2}}+1}\left(1-\frac{\varepsilon_{r1}}{\varepsilon_{r2}}\right)\left[\left(1+\sqrt{\varepsilon_{r2}}\right)u(t) + \left[(1+\gamma)-\sqrt{\varepsilon_{r2}}(1-\gamma)\right]\sum_{n=0}^{\infty}\gamma^n u\left(t-2(n+1)T\right)\right] + \left(\frac{\varepsilon_{r1}}{\varepsilon_{r2}}-1\right)e_x^s u(t)$$

(S33)

Substituting $\gamma = \dfrac{\sqrt{\varepsilon_{r2}}-1}{\sqrt{\varepsilon_{r2}}+1}$ in the previous equation, it is readily to show that $e_x^d(L,t) - e_x^a(L,t) = 0$ for any $t > 0$. Repeating the same procedure for the magnetic field expressions in Eqs. (S23) and (S24), we get $h_x^d(t,L) - h_x^a(t,L) = 0$ for any $t > 0$. This proves that the electric and magnetic fields across the interface of a time-varying dielectric block immersed in a uniform static electric field are continuous.

## 6. Increasing the permittivity of the time-varying dielectric block

In the main text, we have discussed when the initial value of the dielectric block permittivity ($\varepsilon_{r1}$) is suddenly changed in time to a smaller value ($\varepsilon_{r2} < \varepsilon_{r1}$). Here, we discuss the opposite case: the initial value of the dielectric block permittivity is changed in time to a larger value ($\varepsilon_{r2} > \varepsilon_{r1}$). To this end, we assume that the permittivity of the dielectric block of length $L$ changes abruptly from $\varepsilon_{r1} = 4$ to $\varepsilon_{r2} = 8$ at $t = 0$. This change occurs once a uniform static electric field with intensity $e_x^s = 1\text{V}/\text{mm}$ has been established inside the waveguide. With this setup, according to the temporal boundary condition, the value of the electrostatic field in the dielectric region at $t = 0^+$ (right after the permittivity change) is transformed from $1\text{V}/\text{mm}$ to $0.5\text{V}/\text{mm}$ while in the air region it is still equal to $1\text{V}/\text{mm}$. Analogously to the case discussed in the main text, looking at the electric field distribution inside the waveguide at $t = t_1 = 0.1T$ (see Fig. 3S(a)), one can observe a region across the air-dielectric interface where the electric field is continuous and assume a value between $0.5\text{V}/\text{mm}$ and $1\text{V}/\text{mm}$. From to Eqs. (S21) and (S22), we can associate this field with two plane waves ($\frac{1}{2+4\sqrt{2}}u(t_1-(L-z)/c_d)$ and ($-\frac{\sqrt{2}}{1+2\sqrt{2}}u(t_1-(z-L)/c_a)$) propagating away from the section $z=L$ with velocities equal to the phase velocity of the two media. Fig. 3S(b) shows the $z$-component of the instantaneous Poynting vector, $S_z$, at the same time instant ($t=t_1$). $S_z$ is directed along the negative $z$-axis revealing that the energy flows from the air to the dielectric region, conversely to the case with $\varepsilon_{r2}<\varepsilon_{r1}$. This implies that the right-going plane wave ($-\frac{\sqrt{2}}{1+2\sqrt{2}}u(t_1-(z-L)/c_a)$), traveling in the air region, is a backward wave while the left-going plane wave ($\frac{1}{2+4\sqrt{2}}u(t_1-(L-z)/c_d)$), traveling in the dielectric region, is a forward wave. At $t=T$, the left-going plane wave hits the PMC wall generating a second plane wave, $\frac{1}{2+4\sqrt{2}}u(t-(z+L)/c_d)$ (see Eq. (S21)). This wave, adding to the electric field established by the previous wave $\frac{1}{2+4\sqrt{2}}u(t-(L-z)/c_d)$, results in the electric field distribution shown in Fig. 3S(c) at $t=t_2=1.2T$. At this time instant, the air region is still characterized by the same wave $-\frac{\sqrt{2}}{1+2\sqrt{2}}u(t-(z-L)/c_a)$ (see Eq. (S22)) and since $t_2>t_1$, one can observe, as expected, that it has traveled further from the air-dielectric interface. We can notice from Fig. 3S(d), displaying $S_z$ at the same time instant ($t=t_2$), that the flow of energy continues being from the air to the dielectric region. At $t=2T$, the wave traveling in the dielectric region ( $\frac{1}{2+4\sqrt{2}} u\left(t-(z+L)/c_d\right)$) reaches the spatial discontinuity at $z = L$ and generates a reflected wave, $\frac{2\sqrt{2}-1}{18+8\sqrt{2}} u\left(t-(L-z)/c_d - 2T\right)$, and a transmitted wave, $\frac{2\sqrt{2}}{9+4\sqrt{2}} u\left(t-(z-L)/c_a - 2T\right)$, (see Eqs. (S21) and (S22), respectively). These reflected and transmitted waves are added to the electric field established by the previous waves in the dielectric and air regions, respectively, resulting in the distribution at $t = t_3 = 2.2T$ shown in Fig. 3S(e). The energy continues flowing from the air to the dielectric region as shown in Fig. 3S(f). As time passes, the multiple reflection process, due to the spatial boundary at $z = 0$ and $z = L$, continues increasing the number of plane waves propagating into the waveguiding system as described by Eqs. (S21) and (S22). In contrast to the case with $\varepsilon_{r2} < \varepsilon_{r1}$ discussed in the main text, increasing the permittivity of the dielectric block ($\varepsilon_{r2} > \varepsilon_{r1}$), all waves in both regions carry the energy from the air to the dielectric region. It is worth noticing that the value of the electric field in the air region perturbed by the backward waves gets smaller than $1V/mm$ (see Figs. S3(a), (c), and (e)), which is the original electrostatic value. This explains the transfer of the energy from the air to the dielectric region during the transient.

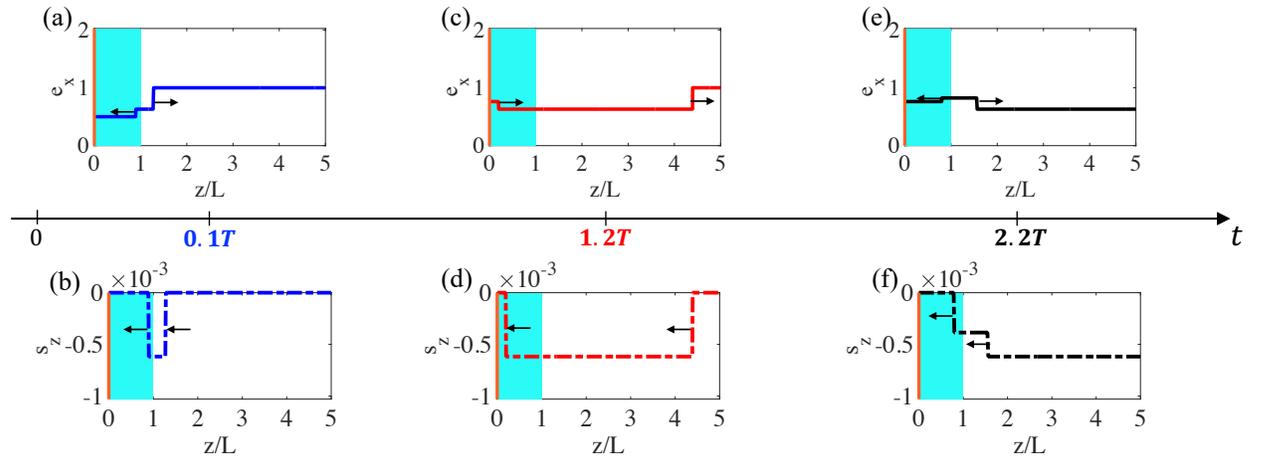

S16

**Figure S3.** Spatial distribution of the electric field, (a), (c), and (e), and the Poynting vector, (b), (d), and (f), at time instants $t = 0.1T$, $1.2T$, and $2.2T$ when the dielectric block relative permittivity changes from $\varepsilon_{r1} = 4$ to $\varepsilon_{r2} = 8$. The arrows in the top panels ((a), (c), and (e)) and in the bottom panels ((b), (d), and (f)) indicate the direction of propagation of the plane waves and of the corresponding Poynting vectors, respectively.

## 7. Total radiated energy density

In this section, we derive and discuss the total radiated energy density carried by the infinite of set of plane waves traveling in the air region. First, we evaluate the instantaneous Poynting vector, $S_z^a(z,t) = e_x^a(z,t) h_y^a(z,t)$, which results

$$S_z^a(z,t) = \frac{(e_x^s)^2}{\mu_0 c_a} \frac{\varepsilon_{r2}}{\left(1-\sqrt{\varepsilon_{r2}}\right)^2} \left(\frac{\varepsilon_{r1}}{\varepsilon_{r2}} - 1\right)^2 \sum_{n=1}^{\infty} \gamma^{2n} \left[ u\left(t - \frac{z-L}{c_a} - 2(n-1)T\right) - u\left(t - \frac{z-L}{c_a} - 2nT\right) \right] +$$
$$+ \frac{(e_x^s)^2}{\mu_0 c_a} \frac{\sqrt{\varepsilon_{r2}}}{1-\sqrt{\varepsilon_{r2}}} \left(1 - \frac{\varepsilon_{r1}}{\varepsilon_{r2}}\right) \sum_{n=1}^{\infty} \gamma^n \left[ u\left(t - \frac{z-L}{c_a} - 2(n-1)T\right) - u\left(t - \frac{z-L}{c_a} - 2nT\right) \right]$$

(S34)

Integrating the previous equation from $t = 0$ to $t \to \infty$, we obtain the total radiated energy per unit area

$$W_{rad} = \frac{1}{2} \frac{\varepsilon_0}{\varepsilon_{r2}} L \left(\varepsilon_{r1}^2 - \varepsilon_{r2}^2\right) \left(\frac{V}{d}\right)^2$$

(S35)

One can notice that $W_{rad} + \Delta W_\infty = 0$, with $\Delta W_\infty$ denoting, as indicated in the main text, the difference of the electrostatic energy per unit area stored in the dielectric block at $t = 0^+$ and $t \to \infty$. The latter equation highlights that the growth (reduction) of the electrostatic energy in the dielectric block induced by the change of its permittivity is fully released (restored) through the electromagnetic energy propagating in the air region.

## 8. Fourier spectrum of the electric field in the air region

The frequency spectrum of the electric field in the air region can be easily obtained by Fourier analyzing $e_x^a(z,t)$ (Eq. (S22)), $E_x^a(z,\omega) = \int_0^\infty e_x^a(z,t)e^{-i\omega t}dt$, which gives

$$E_x^a(z,\omega) = \frac{2e_x^s\sqrt{\varepsilon_{r2}}}{1+\sqrt{\varepsilon_{r2}}}\left(\frac{\varepsilon_{r1}}{\varepsilon_{r2}}-1\right)\frac{\sin(\omega T)}{\omega}\frac{e^{-i\omega\left(\frac{z-L}{c_a}+T\right)}}{1-\gamma e^{-i2\omega T}} + \frac{e_x^s}{i\omega} \quad (S36)$$

with $\omega = 2\pi f$ and $f$ is the frequency. $E_x^a(z,\omega)$, analogously to its time-domain counterpart, is given by the superposition of two terms. The first one,

$E_x^{a,dyn}(z,\omega) = \frac{2e_x^s\sqrt{\varepsilon_{r2}}}{1+\sqrt{\varepsilon_{r2}}}\left(\frac{\varepsilon_{r1}}{\varepsilon_{r2}}-1\right)\frac{\sin(\omega T)}{\omega}\frac{e^{-i\omega\left(\frac{z-L}{c_a}+T\right)}}{1-\gamma e^{-i2\omega T}}$, represents the spectrum of the dynamic field, while the second one, $E_x^{a,dc}(z,\omega) = \frac{e_x^s}{i\omega}$, represents the spectrum of the static field. It is evident from the expression of $E_x^{a,dyn}(z,\omega)$ that its magnitude follows the shape of a *sinc* function and goes to zero whenever $\omega = p\pi/T$ ($f = p/(2T)$) with $p = 1,2,3,\ldots$ Thus, the frequency spectrum of the dynamic electric field in the air region is maximum at $f=0$ and has nulls at equally spaced frequency intervals of $1/(2T)$. Since $T = L/c_d$, the nulls move towards higher frequencies for a short and/or a small $\varepsilon_{r2}$ dielectric block, as shown in Fig. S4.

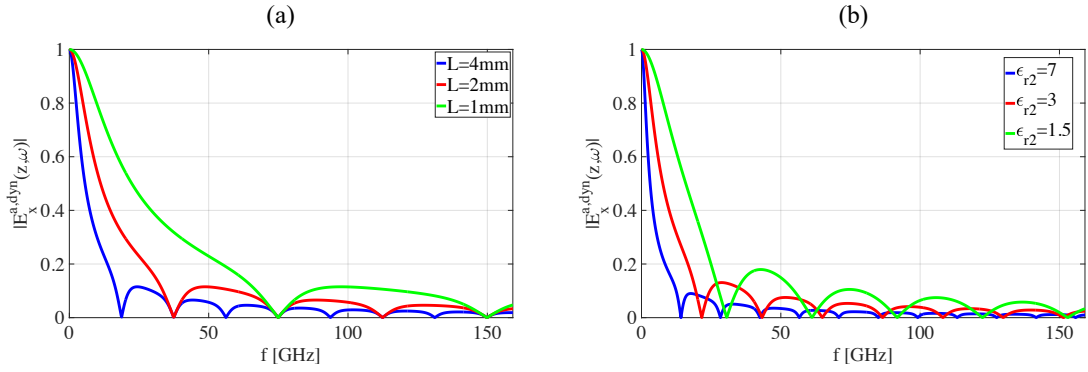

**Figure S4.** Magnitude of the frequency spectrum of the dynamic electric field in the air region normalized to maximum (a) for different values of $L$ ($\varepsilon_{r1} = 8$ and $\varepsilon_{r1} = 4$) and (b) $\varepsilon_{r2}$ ($\varepsilon_{r1} = 8$ and $L = 4\text{mm}$).